%Paper: gr-qc/9509002
%From: Jorge Pullin <pullin@phys.psu.edu>
%Date: Fri, 1 Sep 1995 18:02:31 -0500 (EDT)
%Date (revised): Mon, 4 Sep 1995 10:32:38 -0400

%%%%NOTICE: This newsletter includes a figure using the epsf macro.
%%%%If everything goes wrong, your safest bet is to comment the following
%%%%line and then ignore the tex error that will arise when the figure
%%%%is incorporated. You won't get the figure but you will at least get
%%%%the newsletter,

\input epsf

%%%%%

\magnification=\magstep1
\vbadness=10000
\parskip=\baselineskip
\parindent=10pt
\centerline{\bf MATTERS OF GRAVITY}
\medskip
{The newsletter of the Topical Group in Gravitation of the American Physical
Society}
\medskip
\line{Number 6 \hfill Fall 1995}

\bigskip
\bigskip
\centerline{\bf Table of Contents}
\bigskip
\hbox to 6.5truein{Editorial and Correspondents {\dotfill} 2}
\bigskip
\hbox to 6.5truein{\bf Gravity news:\hfill}
\smallskip
\hbox to 6.5truein{Report from the APS Topical Group in Gravitation,
Beverly Berger {\dotfill} 3}
\smallskip
\hbox to 6.5truein{Some remarks on the passing of S. Chandrasekhar,
Robert Wald {\dotfill} 4}
\bigskip
\hbox to 6.5truein{\bf Research briefs:\hfill}
\smallskip
\hbox to 6.5truein{LIGO Project Status, Syd Meshkov and Stan
Whitcomb{\dotfill} 6}
\hbox to 6.5truein{New Hyperbolic forms of the Einstein Equations, Andrew
Abrahams {\dotfill} 8}
\smallskip
\hbox to 6.5truein{Massless Black Holes, Gary Horowitz{\dotfill} 11}
\smallskip
\hbox to 6.5truein{A plea for theoretical help from the gravity
wave co-op, William  Hamilton  {\dotfill} 13}
\smallskip
\hbox to 6.5truein{Why Quantum Cosmologists are Interested in Quantum
Mechanics, James Hartle {\dotfill} 15}
\smallskip
\hbox to 6.5truein{Probing the Early Universe with the Cosmic
Microwave Background,\hfill}
\hbox to 6.5truein{   Rahul Dave and Paul Steinhardt
{\dotfill} 17}
\smallskip
\hbox to 6.5truein{G measurements, Riley Newman {\dotfill} 21}
\smallskip
\hbox to 6.5truein{Is general relativity about null surfaces?, Carlo Rovelli
{\dotfill} 23}
\bigskip
\hbox to 6.5truein{\bf Conference Reports:\hfill}
\smallskip
\hbox to 6.5truein{ITP program solicitation, James Hartle
{\dotfill} 25}
\hbox to 6.5truein{7th Gregynog workshop in general relativity, Miguel
Alcubierre {\dotfill} 26}
\smallskip
\hbox to 6.5truein{Third Annual Penn State Conference:
Astrophysical Sources of Gravitational Waves,\hfill}
\hbox to 6.5truein{   Curt Cutler  {\dotfill} 28}
\smallskip
\hbox to 6.5truein{General news from GR14, Abhay Ashtekar{\dotfill} 30}
\smallskip
\hbox to 6.5truein{VIth Canadian General Relativity Conference, Jack Gegenberg
{\dotfill} 31}
\bigskip
\bigskip
\bigskip
\bigskip
\leftline{\bf Editor:}

\medskip
\leftline{Jorge Pullin}
\smallskip
\leftline{Center for Gravitational Physics and Geometry}
\leftline{The Pennsylvania State University}
\leftline{University Park, PA 16802-6300}
\smallskip
\leftline{Fax: (814)863-9608}
\leftline{Phone (814)863-9597}
\leftline{Internet: pullin@phys.psu.edu}

\vfill
\eject

\centerline{\bf Editorial}

As you probably noticed from the cover, Matters of Gravity is now the
official newsletter of the APS's TG on Gravitation. More details about
the TG are given in Beverly Berger's article in this newsletter.

{}From a practical point of view little has changed in the operation of the
newsletter. Contributions are still welcome from anyone, the distribution
of the newsletter is free to everyone by email and the correspondents are
still the same, so keep those articles coming! I want to invite especially
those people organizing meetings to send me one or two page reports of what
happened after the meeting. Up to the moment I have tried to keep up with
the growing number of meetings requesting summaries, but my coverage has
been less than adequate.

As usual I wish to thank the correspondents and especially
contributors who made this issue possible. The next newsletter is due
February 1st.

If everything goes well this newsletter should be available in the
gr-qc Los Alamos archives under number gr-qc/9509002. To retrieve it
send email to gr-qc@xxx.lanl.gov (or gr-qc@babbage.sissa.it in Europe)
with Subject: get 9509002 (numbers 2-5 are also available in
gr-qc). All issues are available as postscript or TeX files in the WWW
http://vishnu.nirvana.phys.psu.edu

Or email me. Have fun.

\medskip
Jorge Pullin

\bigskip
\centerline{\bf Correspondents}
\medskip

\parskip=2pt
\item{1.} John Friedman and Kip Thorne: Relativistic Astrophysics,
\item{2.} Jim Hartle: Quantum Cosmology and Related Topics
\item{3.} Gary Horowitz: Interface with Mathematical High Energy Physics,
    including String Theory
\item{4.} Richard Isaacson: News from NSF
\item{5.} Richard Matzner: Numerical Relativity
\item{6.} Abhay Ashtekar and Ted Newman: Mathematical Relativity
\item{7.} Bernie Schutz: News From Europe
\item{8.} Lee Smolin: Quantum Gravity
\item{9.} Cliff Will: Confrontation of Theory with Experiment
\item{10.} Peter Bender: Space Experiments
\item{11.} Riley Newman: Laboratory Experiments
\item{12.} Peter Michelson: Resonant Mass Gravitational Wave Detectors
\item{13.} Stan Whitcomb: LIGO Project
\parskip=\baselineskip

\vfill
\eject

\centerline{\bf Report from the APS Topical Group in Gravitation}
\medskip
\centerline{Beverly K. Berger, Oakland University}
\centerline{berger@oakland.edu}
\bigskip

In April of this year, the Executive Council of the American Physical
Society (APS) approved the formation of the Topical Group in Gravitation.
This action occurred in response to a petition to form such a group signed by
more than 240 APS members (with 200 signatures required). This number
included significant support from physicists whose main area of research is
not gravitation.

The objective of the Topical Group (TG) is to serve as a focus for research
in gravitational physics including experiments and observations related
to the detection and interpretation of gravitational waves, experimental
tests of gravitational theories, computational general relativity,
relativistic astrophysics, solutions to Einstein's equations and their
properties, alternative theories of gravity, classical and quantum cosmology,
and quantum gravity. In order to function effectively to promote this
objective, it is important that the membership of the TG reflect as broad
a cross-section of workers in gravitational physics as is possible.

Signing the petition to form the TG is not equivalent to membership in the
TG. To join, an APS member must check off the appropriate box on the APS dues
payment form. Any APS member who paid dues this year but did not check the
TG in Gravitation box should contact the APS Membership Office ((301)
209-3271, membership@aps.org) to rectify this omission. The Membership Office
should also be contacted by those wishing to join the APS. (APS welcomes
members from all countries.)

Current activities of the TG include planning for two sessions of invited
talks at the APS Spring Meeting (May, 1996 in Indianapolis) and conducting
an election of the Group's first set of officers. The election will
probably occur during October, 1995 with those who have joined the TG
before then eligible to vote.

{\it Matters of Gravity} is now the official
newsletter of the Topical Group in Gravitation and will report on the
group's activities in addition to its broader coverage.

With the support of the community of researchers in gravitation, the TG
can become an effective organization for promotion of the interests of the
field.  All gravitational physicists are encouraged to participate in this
effort!

\vfill\eject

\centerline{\bf  Some Remarks on the Passing of S. Chandrasekhar}
\medskip
\centerline{Robert M. Wald, University of Chicago}
\centerline{rmwa@midway.uchicago.edu}
\medskip

Subrahmanyan Chandrasekhar passed away on August 21, 1995, in Chicago,
at the age of 84, as a result of a heart attack. He had begun to feel
quite ill on the previous night, and, although he managed to drive
himself to the hospital on the morning of August 21, his condition
could not be stabilized after he arrived.

It should not be necessary for me to explain to the readership of this
Newsletter the major role played by Chandra in twentieth century science.
Nor do I feel it appropriate even to attempt to summarize here some of the
highlights of his extraordinary life and scientific career. Fortunately, the
full-length biography by Kameshwar Wali is available to give the reader a
good glimpse into his life. However, I do wish here to mark his passing by
making a few brief, personal remarks.

To those who had met him but did not get to know him well, Chandra must
have seemed an austere and formidable figure. There is some validity to
this impression, since he set the highest standards for himself with regard
to both intellectual rigor and personal integrity, and he was not tolerant of
failings by others in these matters -- though much more tolerant of failings
by others than of his own failings or potential failings. Chandra was
particularly intolerant of scientists motivated primarily by the hope of
receiving recognition by others rather than by a deep, inner conviction
that their work was of importance and interest, whatever anyone else
might think. In order to get to know Chandra, it seemed that a barrier had
to be crossed. I believe that all that was needed to cross this barrier was
some expression to him of one's inner convictions on research or other
intellectual endeavors. It is unfortunate that this barrier had the effect of
isolating Chandra from a portion of the scientific community. Once this
barrier was crossed, the very sensitive, caring, and, above all, loyal nature
of Chandra's personality would become readily apparent. The combination
of these very human qualities of Chandra with his almost super-human
qualities of discipline, self-sacrifice, and dedication to science had a
profound and lasting effect on all those who knew him.

During the last ten years of his life, Chandra's physical stamina declined
noticeably, but his interest in science, his discipline, and his fortitude did
not. His intellectual efforts in writing his recent book on Newton's Principia
were undoubtedly as great as in any of his other major scientific
endeavors. It is very fortunate that he was able to bring these efforts to
completion this winter, less than six months before his passing. Even while
writing this final book, Chandra continued his lifelong pursuit of original
scientific research. In my last full-length conversation with him in early
August, he described what appeared to me to be a very promising
approach to giving a simple derivation of a formula for the gravitational
radiation emitted by a nearly Newtonian pulsating star within the
perturbation framework he had developed with Valeria Ferrari. His
enthusiasm for pursuing this research could not have been very different
than the enthusiasm he must have shown when he had begun doing
scientific research more than 65 years earlier. By chance, Valeria was
passing through Chicago that week on what she thought was going to be a
visit of a few days. Chandra lost no time in convincing her to stay an extra
week to do some calculations to verify the validity of the
approximation he had proposed. It is thus very fitting that Chandra lived
the last weeks of his life pursuing science in much the same manner as he
had done throughout his extraordinary scientific career.

\vfill\eject

\centerline{\bf  LIGO Project Status}
\medskip
\centerline{Sydney Meshkov and Stan Whitcomb, Caltech}
\centerline{syd@ligo.caltech.edu, stan@ligo.caltech.edu}
\bigskip

Since the last {\it Matters of Gravity} report on LIGO in March,
a number of significant events have occurred.  Dedication
ceremonies for the Louisiana LIGO site took place on July 6, 1995
at Livingston, Louisiana, exactly one year after a similar ceremony at
the Hanford, Washington site.  By now the clearing and grubbing
activities at the Louisiana site have been completed and rough grading
will begin shortly.

Working with our Architect/Engineering contractor
(Ralph M. Parsons Co.), LIGO has finalized the conceptual design
for the buildings and associated site development. Parsons has
now started the full design effort. The large scale demonstration
test for the construction of the LIGO beam tubes (which connect
the vertex and ends of the two arms) was successfully completed
this spring, confirming that the design meets our vacuum and
cleanliness requirements. A preliminary design competition for
the remainder of the vacuum system was carried out and a contractor
selected for the final design and fabrication effort.

Organizationally, a LIGO pre-Program Advisory Committee has been
formed, with Peter Saulson(Syracuse) as its chair. The other members
are:  S. Finn(Northwestern), A. Giazotto(Pisa), J. Hall(JILA),
W. Hamilton(LSU), C. Prescott(SLAC), A. Ruediger(MPI-Garching).
This committee will exist only for a year or two. During its brief
life it will act as both a LIGO Program Advisory Committee(PAC)
and as an External Advisory Committee(EAC). Before it goes out
of existence it will help design a final PAC and EAC. The first
meeting is scheduled for September 8-9, 1995 at Caltech.

Following the very successful Aspen Winter Physics Conference on
Gravitational Waves and Their Detection (see {\it Matters of
Gravity}, Number 5, Spring 1995), a second Aspen Winter Physics
Conference has been scheduled for January 15-21, 1996. A major
theme of the Conference will be the study of advanced
interferometers and long range planning. The program will include
extensive meetings of the LIGO Research Community, and several
sessions on LISA.

In the R\&D program, the 40m interferometer at Caltech has been
converted to an optically recombined system, as a first step toward
operating it as a recycled interferometer.  The optical configuration
chosen for the optical recombination is modeled after that planned
for the full-scale LIGO interferometers and uses a small asymmetry in
the arms to produce the required modulation at the point where the
difference in arm lengths is sensed. New servosystems required to hold
the interferometer at its correct operating points are
being testing on the 40 m system, and noise studies to understand the
performance in the new configuration are underway

At MIT, a suspended interferometer to investigate optical sources of
noise at high phase sensitivity is under development. This
interferometer has a simple optical configuration, to emphasize the
study of optical sources of noise and to minimize the amount of time
needed to debug other noise sources.  The initial phase of its
fabrication
has been completed and it is producing its first data, although still
at relatively low power (about 50 mW).  Over the next year, the power
will
be gradually increased, with a goal of achieving shot noise limited
sensitivity with 70 W incident on the beamsplitter.

The effort on the LIGO detectors is growing rapidly as projects move
{}from the R\&D to the actual detector detailed modeling and hardware
design.  A recent highlight in this effort was the integration of
a stabilized argon ion laser under the control of EPICS.  EPICS
is the control system planned for the LIGO detectors, and this
laser is the first detector subsystem to be fully interfaced to it
to provide data logging and operator interfaces appropriate for a
facility the scale of LIGO.

Further information about LIGO can be obtained from our WWW home page
at

{\tt http://www.ligo.caltech.edu}.

\vfill\eject

\centerline{\bf New Hyperbolic forms of the Einstein Equations}
\medskip
\centerline{Andrew Abrahams, University of North Carolina}
\centerline{abrahams@physics.unc.edu}
\bigskip

It is well known that the standard form of the 3+1 Einstein equations
consisting of evolution equations for the 3-metric and extrinsic
curvature, supplemented by prescriptions for determining the
kinematical variables the lapse and shift vector, is not manifestly
hyperbolic.  To produce a d'Alembertian wave equation for the 3-metric
it is necessary to specify the harmonic coordinate condition thus
eliminating the ``undesirable'' second derivatives of the 3-metric
appearing in the Ricci tensor.  The classic hyperbolic systems of
Choquet-Bruhat [1] and Fischer and Marsden [2] are based on this
harmonic coordinate condition.

In the past year there have been four independent attacks aimed at
formulating new hyperbolic versions of general relativity, a common
goal being to relax the restrictions of the 4-dimensional harmonic
condition.  Several of these efforts have been motivated
by the ``Grand Challenge'' effort to simulate the inspiral and
coalescence of black hole binaries.   Besides for the theoretical
advantages for global analysis, it is now widely believed that
manifestly hyperbolic evolution systems will be powerful tools
for numerical solution of the Einstein equations.

The primary obstacle to long-time simulations of black hole spacetimes
is that evolutions using singularity-avoiding foliations (such as
maximal slicing) freeze inside the event horizon leading
to the oft-discussed ``throat stretching''.  Attempting to
numerically resolve the rapidly growing throats for an
astrophysically interesting black hole collision scenario
while preserving reasonable accuracy
is certainly wasteful of computational resources
and currently, at least, beyond our capabilities.  For this
reason, it is now expected that such simulations
will excise the interior of the black holes simply by ignoring
the spacetime inside the apparent horizon (which, assuming cosmic
censorship, lies inside the event horizon).  Since  there
are no general boundary conditions to apply at this surface,
one appeals to causal disconnection and ignores the field inside.
Thus it is clearly critical to have an evolution
system with mathematical characteristics which coincide with
the physical lightcone.  This same feature greatly
simplifies the problems associated with applying {\it outer}
boundary conditions and extracting asymptotic gravitational waveforms;
the fields evolved by the code in the strong-field region can
be directly matched onto radiative variables in the exterior.
Another attractive aspect of such explicitly
hyperbolic evolution schemes is that numerical implementations
can exploit a wealth of sophisticated computational algorithms
developed for the studying the wave equation and the
equations of fluid dynamics.

Fritelli and Reula [3] have developed a first-order symmetric hyperbolic
formulation of general relativity which allows the Newtonian limit to
be taken in a rigorous way.  The basic variables in their
system are the 3-metric, its momentum, and spatial derivatives
of the 3-metric (which obey a new constraint equation).
The lapse and shift vector may be chosen
arbitrarily -- they do not propagate as part of the first-order system.
A somewhat similar approach, also starting with the usual form of the
3+1 Einstein equations has been recently published by
Bona, Masso, Seidel, and Stela [4] following up on earlier work
by Bona and Masso [5].  The basic variables for their system
include the 3-metric, extrinsic curvature, the lapse and shift vector
as well as spatial derivatives of
the lapse, shift, and 3-metric.  They write the evolution system
in flux-conservative first-order symmetric hyperbolic form which
they diagonalize in order to obtain the characteristics.
It should be noted that the kinematical variables have characteristic
speeds in this system which can be different than the speed of light.
Bona et al. have tested their new system with numerical simulations
of Schwarzschild black holes using several different slicing conditions.
They have already seen substantial improvement over calculations using
the standard 3+1 equations.

The systems of van Putten and Eardley [6] and Choquet-Bruhat and York [7,8]
each involve going up one time-derivative from the usual
form of the Einstein equations and producing explicit wave equations
for the dynamical parts of the field.  van Putten and Eardley,
using a tetrad formulation, start from the Bianchi identity and
form Yang Mills-like wave equations for the connection 1-forms.
The Choquet-Bruhat and York system is derived by taking a time-derivative
of the usual evolution equation for the extrinsic curvature
in a manner proposed by Choquet-Bruhat and Ruggeri [9]. (Related
ideas were pursued by Friedrich [10].)
By employing the momentum constraints, a quasi-linear wave equation
for the extrinsic curvature is formed for either harmonic slicing
or slicings where the trace of the extrinsic curvature is specified
in advance.  The main difference between these approaches is
the choice of variables, tetrad vs. standard 3+1.  In both
of these formulations wave equations are produced which are
spatially gauge-covariant.
It is also apparent that the 3-metric evolves but does not, in general,
propagate in a wavelike manner; it does not directly carry the
dynamical degrees of freedom but rather provides an
arena for the wave propagation. This physically intuitive feature
can be seen explicitly in the Choquet-Bruhat/York system by
choosing the harmonic time-slicing, reducing the formally third-order system
to first-order symmetric hyperbolic form, and
reading off the characteristic speeds of the fields.  The 3-metric,
extrinsic curvature, lapse and acceleration all
propagate with zero speed; they are dragged along in a
direction normal to the foliation.  Only the time-dependent tidal forces
have characteristic speeds equal to the speed of light.

It is already clear that these systems will be
fruitful for analytic approximation schemes and
perturbation theory.  As far as numerical
relativity is concerned, formal and aesthetic
differences may not be adequate to identify the
best scheme.  The proof of successful numerical
implementation will be in the
non-linear gravitational pudding resulting from the
violent coalescence of two black holes.

{\bf References}

\noindent [1] Y. Choquet-Bruhat, Acta. Math. {\bf 88}, 141 (1952).

\noindent [2] A. Fischer and J. Marsden, Comm. Math. Phys. {\bf 28}, 1 (1972).

\noindent [3] S. Fritelli and O. Reula, Comm. Math. Phys. {\bf 266},
221-245 (1994).

\noindent [4] C. Bona, J. Masso, E. Seidel, and J. Stela, Phys. Rev. Lett.,
{\bf 75}, 600 (1995)

\noindent [5] C. Bona and J. Masso, Phys. Rev. Lett., {\bf 68}, 1097 (1992).

\noindent [6] M.H.P.M. van Putten and D.M. Eardley (gr-qc/9505023, 1995).

\noindent [7] Y. Choquet-Bruhat and J.W. York, C.~R.~Acad.~Sci. Paris,
in press
(gr-qc/9506071, 1995).

\noindent [8]A. Abrahams, A. Anderson, Y. Choquet-Bruhat, and J.W. York Jr.,
(gr-qc/9506072, 1995).

\noindent [9] Y. Choquet-Bruhat and T. Ruggeri, Comm. Math. Phys. {\bf 89},
269 (1983).

\noindent [10] see, e.g., H. Friedrich, Comm. Math. Phys., {\bf 91} 445,
(1983).

\vfill\eject

\centerline{\bf Massless Black Holes}
\medskip
\centerline{Gary T. Horowitz, University of California at Santa Barbara}
\centerline{gary@cosmic2.physics.ucsb.edu}
\bigskip

 The idea of a massless black hole almost sounds like a contradiction in
 terms. A black hole is supposed to be a massive object which has
 undergone gravitational collapse. What could it mean for a black hole
 to be massless?
 Surprisingly, it has recently been shown that `massless black holes' play
 an important role in string theory. They have appeared in at least two
 different contexts, and  in each case were instrumental in resolving
 long standing problems. The first problem concerned the
 strong coupling limit of
 string theory in ten dimensions, and the second involved some troubling
 singularities in the low energy description of string theory in four
 dimensions.

One example of a massless black hole was discussed by Witten [1],
who made the following
argument (see also [2]).
We usually think of an extremal black hole as having a mass equal to its
charge $M = |Q|$. However, if one is careful about coupling constants,
one finds that (for a certain ten dimensional string theory called Type IIA)
the relation is really $$M = |Q|/g$$ where $g$ is the coupling constant.
These extremal black holes are supersymmetric, and one can use this fact to
argue that this formula does not
receive any quantum corrections even
in the limit of strong coupling. In the extreme limit where $g \rightarrow
\infty$, the mass of these extreme black holes goes to zero. This suggests
that these black hole solitons become light degrees of freedom
in the strong coupling limit of the theory. In fact, if one assumes that
the charge is quantized and that different charges correspond to distinct
states, one is lead to a tower of states with $M = c|n|$ for some small
constant
$c$. This is exactly the same spectrum that  one obtains from Kaluza-Klein
compactification of one dimension on a circle. As the radius of the circle
becomes large, the massive Kaluza-Klein states become light just like
the black holes at strong coupling. Using this connection, it has
been conjectured that the low energy dynamics of the
strongly coupled ten dimensional string theory is, in fact, {\it eleven
dimensional} supergravity.

A second example of  massless black holes in string theory was provided
by Strominger [2]. A standard method of reducing string theory to
four dimensions is to compactify it on a six dimensional
manifold called a Calabi-Yau space. However, it has been known for
some time that this approach suffers from the following difficulty.
There are continuous deformations of the Calabi-Yau space which give
rise to scalar
`moduli' fields in four dimensions. In particular, there is a deformation
which corresponds to a three-surface in the Calabi-Yau space shrinking
down to zero size. At this point, the effective four dimensional
lagrangian becomes singular, since the kinetic energy term for one
of the moduli fields acquires a diverging coefficient.
(Notice that this is a singularity in the
action itself, and not a particular solution.) Strominger resolved this
singularity as follows. The ten dimensional theory contains three dimensional
extended objects surrounded by an event horizon [4] . One can think of them
as `black three-branes'. These objects carry a mass per unit three-volume
$M/V$
and charge $Q$, and there is
an extremal limit in which $M = |Q| V $.
Now, when one compactifies on a Calabi-Yau space, this
black three-brane can wrap around a nontrivial three-surface in the compact
space. From the four dimensional viewpoint, the three-brane now
looks like an ordinary black hole. However, when the volume of the
three-surface
shrinks to zero, the mass of this black hole goes to zero. Near this point,
the black hole can no longer be treated classically, but instead
acts like another light `particle' which should be included
in the low energy lagrangian.
Strominger has
shown that the singularity in the standard four dimensional action is exactly
what one would get if one views this as an effective action obtained by
starting with a nonsingular action with one additional
massless charged scalar field, and integrating it out.
Perhaps even more striking is the fact [5]
that one can  use these massless black holes
to give an apparently nonsingular description of  a transition from one
Calabi-Yau space to a topologically different one.

To summarize, in the above examples, extremal black holes act just like
elementary particles (or in string theory, just like other states of the
string). The main difference is that they can carry certain charges which are
not carried by ordinary particles.
This clearly suggests that there may be no fundamental distinction between
extremal black holes and elementary particles.

{\bf References}

\noindent [1]  E. Witten, Nucl. Phys., {\bf B443}, 85, (1995).

\noindent [2] C. Hull and P. Townsend, Nucl. Phys. {\bf B438}, 109, 1995;
P. Townsend, Phys. Lett. {\bf B350}, 184, (1995).

\noindent [3] A. Strominger, ``Massless Black Holes and Conifolds in
String Theory", hep-th/9504090.

\noindent [4] G. Horowitz and A. Strominger,  Nucl. Phys.,
{\bf B360}, 197, (1991).

\noindent [5] B. Greene, D. Morrison, and A. Strominger, ``Black Hole
Condensation and the Unification of String Vacua'', hep-th/9504145.

\vfill\eject
\centerline{\bf  A plea for theoretical help from the gravity wave co-op}
\medskip
\centerline{Bill Hamilton, Louisiana State University}
\centerline{hamilton@phgrav.phys.lsu.edu}
\bigskip

     With the most recent successes in operating cryogenic
gravity wave detectors at LSU, Rome and Perth and with the recent
discovery of the TIGA configuration for an omnidirectional
detector we see very great advantages to operating resonant TIGA
detectors in coincidence with the upcoming LIGO and VIRGO
detectors. Carl Zhou's paper in the last PRD shows that such a
configuration will be very powerful in determining a source's
location in the sky. The gravity wave co-op has started to design
TIGA detectors to run in coincidence with the interferometers. As
shown by Johnson and Merkowitz we expect such detectors to be
more sensitive than the interferometers in their higher and
narrower frequency range. There are problems, however.

     The gravity wave co-op has been told the only way that a new
detector can possibly be built is if there is a guaranteed source
for it to go after. It probably is not sufficient, we are told,
to depend on supernova sources since most recent models have a
much softer collapse and much less GW emission than we expected
several years ago.

     Thomas Stevenson at the University of Maryland did an
estimate of the detectability of a NS-NS inspiral and merger,
based on the recent work of Centrella et al. His conclusion was
that, for a source at 15 Mpc, the signal to noise ratio in a
nearly quantum limited TIGA would be about 4 in the largest TIGA
that we can consider practical, a 60 tonne detector at 870 Hz.
That's too low for a detection with only one detector and even
too low for a two detector coincidence. Most of the energy that
would be detected would come from the final few revolutions of
the inspiral with very little coming from the actual merger.
Warren Johnson and I had separately done a very cursory look at
Centrella's results and came to a similar discouraging
conclusion.

     On the other hand, there are those who argue that
Centrella's work is just the first cut in a very difficult
problem and that a different model for the star could give vastly
different results. Williams and Tohline, for instance, used a
fluid model code to investigate inspiraling normal binary stars.
They found a bar like central region after the merger which
resulted in a very large time dependent quadrupole moment.
Tohline suspects that his code would give a similar result in the
NS-NS case. Some of us naively hope that the formation of rapidly
spinning black holes, perhaps at the end of a merger, will prove
to be an efficient source.

     Those of us who build detectors are simple plumbers who
don't have access to the latest information about sources or the
insight to understand what is good and what is bad about various
numerical approaches to the problem. We need good input about
possible sources, the type of statistics that we might expect
{}from these sources, and the character of the gravitational wave
signal that might be expected. The frequencies of interest to the
resonant detectors are those above 1 kHz. The detectors that we
know how to build will probably be pretty noisy above 3 or 4 kHz
so that is probably the upper limit of what we should consider
building.

     This letter then is an urgent request for any input about
possible high frequency sources. We need to have a good plausible
source to look for with reasonable statistics. We need to get the
search underway as soon as possible because it will take at least
a couple of years to design and build the detector and it will
probably take some time to work out the details of the sources
that we think we can detect.

     With LIGO now well underway some may believe that its great
sensitivity will eliminate the need for resonant burst detectors.
Our experience is that confidence in any detection will be
enormously enhanced if there are more detectors involved. It will
certainly be more convincing if a gravitational wave detection is
made with one technology and confirmed by another.

     To build the omni-directional TIGA we need sources. We
cannot just improve technology in the hope that we will see
something. Please give serious thought to possible high frequency
gravitational wave sources. It is important for all of us if we
are to begin to do gravitational wave astronomy.

\vfill\eject
\centerline{\bf Why Quantum Cosmologists are Interested in Quantum
Mechanics}
\medskip
\centerline{James B. Hartle, University of California at Santa Barbara}
\centerline{hartle@cosmic.physics.ucsb.edu}
\bigskip

The objectives of quantum cosmology are a description of the universe
in quantum mechanical terms, and, within that description, a theory of
the universe's initial condition that makes testable predictions about
observations today. Early work on this subject concentrated on the
predictions of particular theories of the initial condition using
semiclassical approximations to quantum mechanics.  More recently
there has been interest on the part of some quantum cosmologists, and
others in quantum gravity, in generalizing the usual framework of
quantum mechanics and in clarifying its interpretation when applied to
closed systems such as the universe.  A somewhat representative (but
not exhaustive!) list of recent papers in this vein is given below.

The first reason that quantum cosmologists are interested in quantum
mechanics is that a generalization of the usual theory may be needed
to deal with quantum spacetime geometry. A fixed spacetime geometry is
central to the usual formulations of quantum mechanics, for instance,
to give meaning to the time which enters so basically. But, in a
quantum theory of gravity, spacetime geometry is not fixed. It is
rather a quantum dynamical variable, fluctuating and without definite
value. This conflict is called the ``problem of time'' in quantum
gravity. (For lucid reviews see refs [9,10].) A long investigated route
to a resolution of this question has been to keep quantum mechanics as
is, but find preferred time(s) in general relativity
[9,10,11]. Recently, however, there has been interest in the opposite
contrast -- keep general relativity as is, but generalize quantum
theory, so that it does not require a fixed background spacetime, but
yields the usual theory in situations where spacetime geometry is
approximately fixed and can supply a fixed notion of time [12,13].

A second reason that quantum cosmologists are interested in quantum
mechanics is that, in applying quantum mechanics to the universe as a
whole, the interpretive difficulties of the subject are encountered in
an unavoidable manner. To utilize the probabilities of density
fluctuations at the time the observed universe was the size of an
elementary particle one needs to be clear about what these
probabilities mean!  Phenomena such as classical behavior, that in the
familiar frameworks might be posited or taken for granted, now require
explanation in a wholly quantum universe.  The range of ideas that
have been put forward for quantum frameworks for the universe as a
whole are well beyond what could be described or critically analyzed
in this small compass. One contrast that emerges is between those who
would put observers, measurement, and even consciousness in a central
position in quantum theory [1,2], and those who, with the aim of
precision, would expel these elements from so fundamental a role
[3-8].

How are we to evaluate generalizations of quantum mechanics that have
been proposed?  How are we to distinguish between the various
interpretive ideas? The methods for doing this are standard in
physics. In cosmology we may ask: Are the proposed frameworks general
enough to calculate the probabilities for alternative behaviors of the
universe -- its large scale behavior and that of everything inside it
including quantum spacetime? Are the frameworks logically consistent
and conceptually and mathematically precise? Do they allow the
assumptions of quantum theory to be stated more precisely and clarify
long standing issues of interpretation? Do they lead to calculations
of observable quantities that can be carried out in the near term? Do
they correctly and easily reproduce what is already well tested in
quantum mechanics and field theory? Do they differ one from another in
testable predictions? Do they further the research program of
understanding quantum gravity and the quantum origin of the universe?

More important than the differences between the ideas under discussion
is the common thread that links them. By focusing on a program of
predicting the observable features of this largest of physical systems
{}from a theory of its quantum initial condition, we may be led to a
clearer and more general form of quantum theory -- perhaps even with
implications for the laboratory.

\parskip=5pt
{\bf References}

\noindent [1] D. Page,  quant-ph/9506010.

\noindent [2] C. Rovelli, hep-th/9403015.

\noindent [3] H.F. Dowker and A.~Kent,  gr-qc/9412067.

\noindent [4] L.Diosi, N. Gisin, J.J. Halliwell and I.G. Percival,
 {\sl Phys.Rev.Lett} 74, 203 (1995).

\noindent [5] C. Kiefer and H.D.~Zeh, {\sl Phys. Rev. D}, 51, 4145 (1995).

\noindent [6] R.D. Sorkin, gr-qc/9401003.

\noindent [7] J.P.~Paz and W.H.~Zurek, {\sl Phys. Rev. D}, 48, 2728, (1993).

\noindent [8] M.~Gell-Mann and J.B.~Hartle in {\sl Complexity, Entropy,
and the Physics of Information}, ed. by W. Zurek,  Addison Wesley, Reading, MA
(1990).

\noindent [9] K.~Kucha\v r, in {\sl Proc. 4th Canadian
Conference on General Relativity and Relativistic Astrophysics}, ed.~by
G.~Kunstatter, et. al. World Scientific, Singapore,
(1992).

\noindent [10] C.~Isham, in {\sl Integrable Systems, Quantum Groups,
and Quantum Field Theories}, ed by L.A.~Ibort and M.A.~
Rodriguez, Kluwer Academic Publishers, London (1993).

\noindent [11] W.~Unruh and R.~Wald,  {\sl Phys. Rev. D} , 40, 2598 (1989).

\noindent [12]  C.J.~Isham, {\sl J. Math. Phys.}, 35, 2157 (1994).

\noindent [13] J.B.~Hartle, in {\sl Gravitation and Quantizations}
(Les Houches '92),
edited by B. Julia and J. Zinn-Justin, (North Holland, Amsterdam, 1995),
gr-qc/9304006.

\vfill\eject
\centerline{\bf
 Probing the Early Universe with the Cosmic Microwave Background}
\medskip
\centerline{Rahul Dave and Paul Steinhardt, University of Pennsylvania}
\centerline{dave@steinhardt.hep.upenn.edu,steinh@steinhardt.hep.upenn.edu}
\bigskip

The cosmic microwave background (CMB), first discovered by Penzias and
 Wilson in 1964, provides a critical test of the big bang model and
 provocative ideas that go beyond the big bang to explain the origin
 and evolution of large scale structure in the universe [1]
 The CMB may also be used to constrain the values of cosmological
 parameters such as the Hubble expansion rate and the matter density
 of the universe.

According to the big bang model, the CMB is radiation emitted some
 100,000 years after the big bang.  Prior to that time, the universe
 consisted of a hot, dense gas of free electrons and nuclei in
 equilibrium with photons.  After 100,000 years or so, the universe
 had cooled enough for the free electrons and nuclei to combine into
 neutral atoms. From that time onwards, the photons, which interacted
 only very weakly with the newly neutralized medium, began to freely
 stream in all directions.  Initially, the spectrum was perfectly
 black body with a temperature of nearly 10,000~K.  Over the
 subsequent 10 billion years, the photon distribution red shifted due
 to the expansion of the universe.  The spectrum remains black body
 today but now the average energy lies in the microwave regime.

Recently, the Far Infra-Red Absolute Spectrophotometer (FIRAS)
experiment on board the COsmic Background Explorer (COBE) satellite
measured the spectrum of the radiation and confirmed the big bang
predictions.  With exquisite experimental precision, the spectrum was
shown to be perfect black-body with a temperature of $T_0
=2.726\pm0.010$~K [2].

The focus of cosmologists has recently turned to what can be learned
{}from the CMB anisotropy, the difference in blackbody temperature along
different lines of sight on the sky.  The anisotropy can be used to
test ideas about the very early evolution of the universe and the
origin of large-scale structure.  It was nearly 30 years after the
discovery of the CMB before any non-uniformity in the CMB temperature
across the sky was detected. The temperature variation is so tiny that
instruments with microKelvin sensitivity had to be developed. The
first successful detection was by the Differential Microwave
Radiometer experiment aboard the COBE satellite (1992) [3].

The root-mean-square fluctuation in temperature, found to be roughly
$0.001\%$, is an important cosmological parameter for understanding
the formation of galaxies.  How did the universe evolved from being
highly homogeneous at the 100,000 year mark, as imaged by the CMB, to
being highly inhomogeneous today, as shown in recent maps of the
distribution of galaxies?  The favored explanation has been
``gravitational instability'', the amplification of inhomogeneity
caused by gravity drawing additional matter into overdense regions and
away from underdense ones.  A straight-forward calculation shows that
the inhomogeneity must have been $~0.001\%/\Omega$ at the 100,000 year
mark in order for gravitational instability alone to explain the
inhomogeneity seen today.  Here $\Omega$ is the ratio of total energy
density of the universe to the critical density, the threshold that
separates an open, ever-expanding universe from a closed universe that
ultimately recollapses.  The COBE observation has provided important
support for the gravitational instability concept provided $\Omega$ is
not much smaller than one.

The key, unresolved issue is: What created the anisotropy in the first
place? A leading explanation is the inflationary [4]  model of
the universe.  In this model, the seeds of large-scale structure are
quantum fluctuations in the energy density generated when the
observable universe occupied a sub-nucleonic size just instants after
the big bang.  There then followed a brief burst of extraordinary
superluminal expansion ({\it inflation}) in which the microscopic
quantum fluctuations were stretched into a spectrum of energy density
perturbations that span cosmic dimensions.

The fluctuation spectrum is predicted to be nearly scale-invariant.
If one expands the energy density field $\rho({\bf x)}$ in a sum of
fourier modes with amplitude $\delta(\lambda)$, then the amplitude of
a mode is nearly independent its wavelength $\lambda$.  If the
spectrum is parameterized by a spectral index, $n$, defined by
$\delta(\lambda) \sim \lambda^{(1-n)/2}$, then a precisely
scale-invariant spectrum corresponds to $n=1$ (See [1] for
more precise definitions and discussion).  In inflationary models this
index can be in the range $0.7 \le n \le 1.2$.

Given a map showing the temperature variation of the CMB across the
sky, the best test of inflation or competing models is the temperature
auto-correlation function, defined as
$$  C(\alpha)  \equiv  {\langle {\delta T \over T_0}({\bf x})
{\delta T\over T_0}({\bf x'}) \rangle}_{{\bf x}.{\bf x'}=\cos \alpha}
\equiv   \sum_{l} {2l+1\over4\pi} C_l P_l (\cos \alpha)$$
where $\langle\ldots\rangle$ is an all-sky average over every pair of
directions separated by anlge $\alpha$.  The values of the $C_l$'s,
called {\it multipole moments}, depend on the average temperature
variation between two directions separated by $100/l$ degrees ($\pi/l$
radians). A plot of $l (l+1) C_l$ vs. $l$ is called the CMB power
spectrum (see Fig.~1, which is normalized by the COBE value of the
power spectrum at $l=9$).

The power spectrum is a tell-tale fingerprint that can be used to
distinguish competing cosmological models.  The characteristic power
spectrum predicted by inflation (Fig.~1) has a plateau for $l < 100$
(large angular scales) and a series of peaks for $l>100$.  Variations
in the energy density result in varying gravitational potentials that
red shift or blue shift the CMB photons by different amounts across
the sky, producing apparent CMB temperature differences across the
sky.  If the energy density fluctuations are scale-invariant, the
variations they induce on the CMB temperature are too, resulting in a
power spectrum independent of $l$.  This is true for those energy
density fluctuations that have not evolved since inflation, which are
the fourier modes relevant to $l<100$.  These modes have wavelengths
$\lambda \gg 100,000$~light-years, and so no redistribution of matter
could have happened across them by the time the CMB radiation first
began free-streaming.

The  COBE experiment, which was sensitive only to the long-wavelength
modes affecting multipoles with  $l<30$, constrained
the spectral index $n$ to be $0.91\pm0.36$ [5].
This value is consistent with the inflationary prediction, although
some other models of large-scale
structure formation, such as cosmic strings and textures, make a similar
prediction. As shown in Fig.~1, i
ndependent experiments sensitive to $l<100$ have
yielded consistent results.

\medskip\epsfxsize=320pt \epsfbox{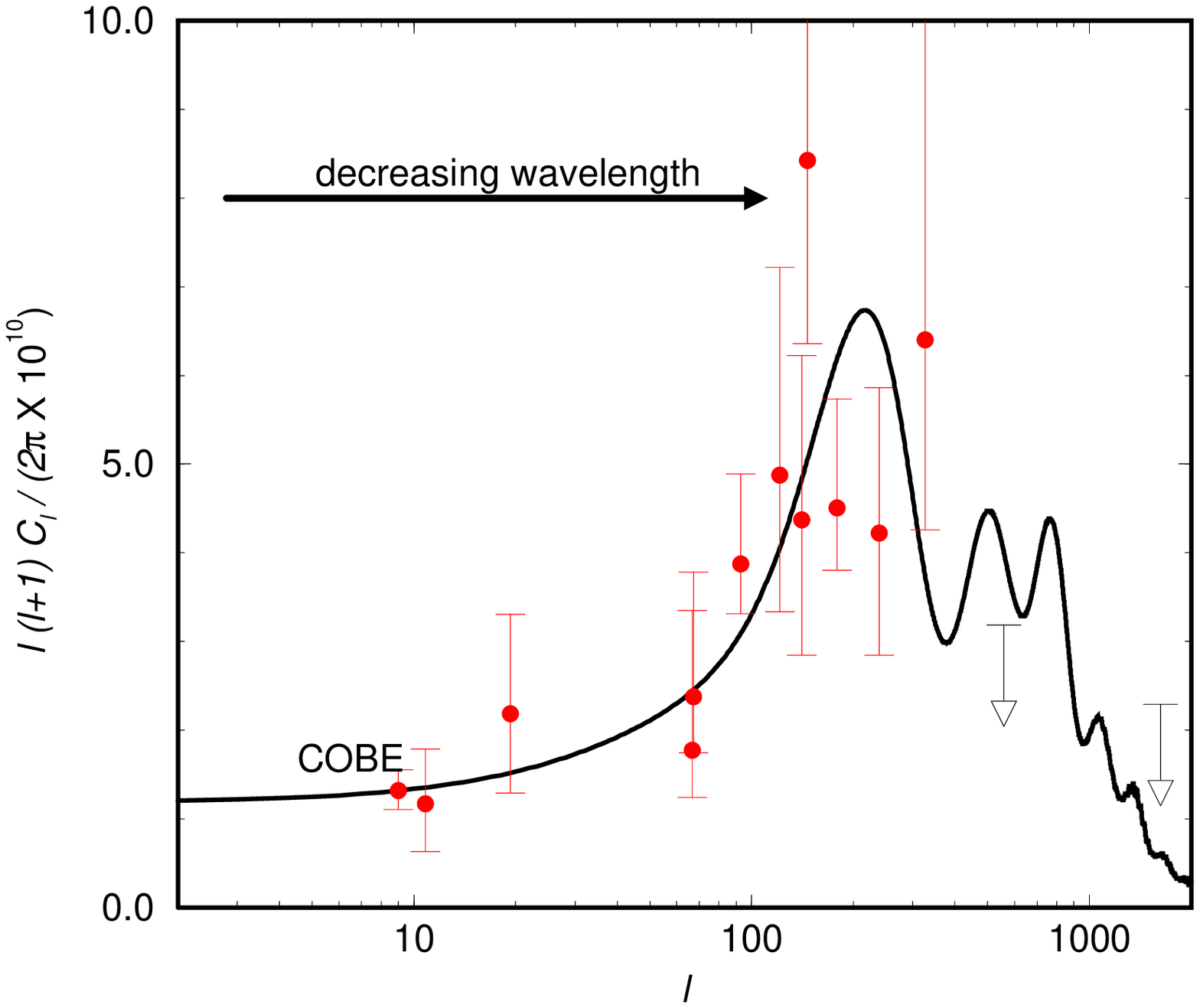} {\bf Figure 1}
{\it A plot of the CMB
power spectrum vs. multipoles for a flat universe with Hubble
expansion rate 50 km/s/Mpc, and composition 5\% baryons with the rest
Cold Dark Matter. The power spectrum is normalized to COBE. Results
{}from experiments are overlaid.  As we move from left to right in $l$,
we probe anisotropies created by fluctuations of decreasing
wavelength. The key features (described in the text) are the plateau
for $l < 100$ and the oscillations beginning at $l \sim 220$.}

The behavior for $l>100$ is different because the energy density
fluctuation  modes that perturb the CMB have wavelengths much shorter
than 100,000~light-years.
Matter and radiation have had time to redistribute across the wavelength
before the CMB was emitted. In particular, there are
 a series of acoustic oscillations
in which the matter and photons are drawn together by gravity and then bounce
back due to the radiation pressure.  The effect of the acoustic oscillations
on the CMB photons is to produce the series of peaks shown in the figure.

The discovery of CMB anisotropy peaks would be extremely significant since it
would be an unmistakable verification of inflation.
The location of the left-most peak along the $l$-axis
is a sensitive test of $\Omega$; a peak at $l\sim 200$ is
support for the inflationary predictions $\Omega=1$.
This method for measuring $\Omega$ is extremely
powerful compared to present methods because
it probes the universe over greater distances and counts all
 forms of energy density including
baryonic matter, dark matter, unclustered matter and
cosmological constant.

The detailed shape of the first and subsequent peaks depend
sensitively on cosmological parameters.  Precise measurements,
especially when combined with more conventional astronomical
observations, will determine the energy densities in baryons and
various species of dark matter, the vacuum energy density or
cosmological constant, and the Hubble expansion rate.

A number of ongoing experiments are attempting to measure the CMB
anisotropy with half-degree resolution or better to determine the
existence and shape of any peaks.  Whereas the COBE experiment is
aboard a space platform, the others are performed at remote, high, and
dry locations such as Saskatoon and the South Pole or are launched in
high-altitude balloons.  Fig.~1 shows the present experimental
situation. It can be seen that, while the error bars on present
experiments are too large to make the results conclusive, there seems
to be reasonable agreement with the inflationary predictions.

There have been rapid improvements in observational strategy and
technology.  The detectors on most experiments launched today are more
than an order of magnitude more sensitive than those aboard COBE.
Several groups are now pioneering long-duration balloon experiments in
which a balloon-based apparatus circumnavigates Antartica for several
weeks. There are also an international efforts underway to launch a
second satellite experiment with high angular resolution
instruments. There is a much improved understanding of how to remove
foreground signals from the data and how to use the CMB anisotropy to
test cosmological models.  Thus, there is every hope that a precise,
high resolution map of CMB temperature variations across the sky will
be available within the next decade and that this snapshot of the
early universe will be a historic contribution to our understanding of
the origin and evolution of the universe.

{\bf References:}

\noindent [1] P. J. Steinhardt,
{\it IJMPA}  {\bf A}10, 1091-1124 (1995) reviews
inflation and its effects on the microwave background.
An expanded version of this paper,
entitled ``Cosmology at the Crossroads," updates of experimental results, and
software tools for analyzing the CMB can be found on the internet at
 http://dept.physics.upenn.edu/~www/astro-cosmo/.

\noindent [2] J. C. Mather, {\it et al.\ }, {\it Ap.\ J.\ }{\bf 420},
439 (1994).

\noindent [3] G. F. Smoot {\it et al.\ }, {\it Ap.\ J\ }{\bf 396}, L1 (1992).

\noindent [4] A. H. Guth and P. J. Steinhardt, "The Inflationary Universe" in
{\bf The New Physics}, ed. by P. Davies, (Cambridge U. Press, Cambridge,
1989) pp. 36-60.

\noindent [5] K. M. Gorski, {\it et al.\ }, {\it Ap.\  J.\ }{\bf 430},
L89 (1994).

\vfill\eject

\centerline{\bf G Measurements}
\medskip
\centerline{Riley Newman,
University of California, Irvine}
\centerline{rdnewman@uci.edu}
\bigskip

  The highly discrepant results of G measurements which I reviewed
in Newsletters 3 and 4 have attracted public attention upon their
presentation at the spring APS meeting in Washington this year.
Values of G presented there by the German PTB group and by the New
Zealand group were those reported earlier (Newsletter \#4).  The
group at Wuppertal University in Germany had previously reported a
G with 117 ppm accuracy in good agreement with the CODATA value
(0.6$\sigma$), but now report 6.6685 ($\pm$ 140 ppm), 3.2$\sigma$
below CODATA: 6.6726 ($\pm$ 128 ppm), and 2.6$\sigma$ above New
Zealand: 6.6656 ($\pm$ 95 ppm).  The big puzzle remains the PTB
result: 6.7154 ($\pm$ 83 ppm), 42$\sigma$ above CODATA and even
further from the other new values.  This result, the product of
many years of carefully repeated measurements by the German NIST
equivalent laboratory, is not to be lightly dismissed.

  Gabe Luther continues his G measurement at Los Alamos.  I have
lost email contact with the Russian group that has been measuring
G.  Not surprisingly, several new G measurements are under way or
planned; I am aware of the following:

  Jim Faller's group at JILA is using its free fall g measurement
instruments with a movable local mass to measure G, initially at a
level of about a percent to explore the feasibility of a more
accurate measurement.   Jens Gundlach of the University of
Washington \lq\lq E\"{o}tWash'' group plans a measurement using
their rotating torsion balance in a novel fashion:  the turntable
rotation speed will be servoed to null the sin(2$\theta$)
acceleration signal from a torsion pendulum due to a pair of
attracting masses.  The measured acceleration of the turntable
then reflects the acceleration of the pendulum, while the fiber
never twists.  This nicely avoids troublesome issues of fiber
deflection calibration and possible fiber nonlinearities.  The
pendulum is to be a thin plate suspended vertically; thus its
quadrupole moment (which largely determines the gravitational
torque) is nearly proportional to its moment of inertia I.  The
measured angular acceleration which determines G is then highly
insensitive to the mass distribution and geometry of the pendulum.
By also rotating the source masses and averaging over turntable
orientations, the effects of turntable readout nonlinearities and
ambient fixed gravitational fields are averaged out.  Jens finds
measured turntable acceleration noise levels should allow a 100
ppm G determination in less than a day.

  At UCI we plan a G measurement using a cryogenic torsion
pendulum, using the now classic method of measuring the pendulum's
oscillation frequency as a function of source mass orientation.
We too will use a thin vertical plate for the pendulum.  The
source masses will be a pair of rings positioned so that their
multipole couplings to the pendulum ideally vanish for $\ell =
1,3,4, $ and 5, leaving an essentially pure quadrupole coupling.
The pendulum will oscillate at one of five amplitudes between 2
and 9 radians at which the frequency shift as a function of
amplitude is an extremum and hence insensitive to error in
amplitude determination.  Large amplitude operation raises
concerns of fiber nonlinear effects, which have often been viewed
as a possible source of systematic error in such experiments;
however our measurements of nonlinearities in an aluminum 5056
fiber at 4.2K, as reflected in the harmonic content and frequency
amplitude dependence of a symmetric pendulum operating at large
amplitudes, indicate that fiber nonlinearities should affect a G
measurement at a (correctable) level of only a few ppm.
Comparison of G values determined at a variety of oscillation
amplitudes will provide a powerful consistency check for
systematic effects.  We aim for 20-50 ppm accuracy in an initial
experiment, with a goal of 1-5 ppm accuracy in a second phase
using fused silica source mass rings.

  At the APS meeting, Craig Spaniol (West Virginia State College)
and John Sutton (Goddard Space Flight Center) announced their hope
to organize a multi-institutional program for a new G measurement.
Spaniol may be contacted at 304-766-4123.

  Several proposals have been made for G measurements in space
(outside my province as reporter on ``laboratory gravitation
experiments'').  A discussion of several of these proposals
appears in a paper to be published by Alvin Sanders and George
Gillies.

\vfill\eject

\centerline{\bf Is general relativity about null surfaces ?}
\medskip
\centerline{Carlo Rovelli, University of Pittsburgh}
\centerline{rovelli@phyast.pitt.edu}
\bigskip

What has Ted Newman been doing during the last decade?  Some papers of
him appeared, but too few for Ted's volcanic creativity.   Indeed, for over a
decade Ted Newman's main activity has been to pursue a strange and
ambitious program with by a small band of faithful collaborators, primarily
Carlos Kozameh and Simonetta Frittelli.   Now the three are emerging from
the long search with a stream of papers presenting a surprising new rabbit
in the hat: a reformulation of general relativity as a theory of evolving
surfaces, whose dynamics is determined by an equation in six dimensions.

The group has gone through a long sequence of twists and shifts, during
which it has been talking of light-cone cuts on Scri, holonomies of the self-
dual connection along light rays, sphere-worth sets of coordinate
transformations, and so on.  Most of this language has now been left behind,
and the formulation of GR finally discovered is simple and tidy; and
surprisingly different from the way we use to think of GR or any other field
theory.

Let me try to summarize this reformulation, at the cost of much vagueness
and much simplification.  Consider an open region $M$ of a four dimensional
manifold, and a foliation $s$ of $M$. (A foliation is a family of
non-intersecting surfaces filling $M$. The surfaces are determined as level
surfaces of a suitably regular function $s: M\rightarrow R$).  Clearly, there
exists (locally) a metric tensor $g$ on $M$ --determined up to a conformal
rescaling-- such that each  surface of the foliation $s$ is null.  Let us next
consider several (overlying) foliations of $M$, or, more precisely, a
one-(complex-)parameter family $s(\zeta)$ of such foliations; $\zeta$ is a
complex parameter.  In general, there will not be a metric $g$ such that
each surface in each foliation is null.   Frittelli, Kozameh and Newman have
found the conditions on the family of foliations $s(\zeta)$, such that $g$
exist, plus an extra equation for the undetermined conformal factor which
implies that $g$ is an Einstein metric.

Since the theory is generally covariant, the physical content of the family
of foliation s($\zeta$) is given by the position of the overlying foliations
with respect to each others.  This fact can be used as follows. The family of
foliations s($\zeta$) determines a preferred coordinate system x for every
z --essentially by posing $x^1=s$, $x^2={\partial s\over\partial\zeta}$,
$x^3={\partial s\over\partial\bar\zeta}$, $x^4= {\partial^2 s \over\partial
\zeta \partial\bar\zeta}$.  The other second derivatives, namely $\Lambda=
{\partial^2 s \over\partial\zeta \partial\zeta}$ contain physical
information. These can be expressed in the preferred coordinate system as
$\Lambda(x;\zeta)$. Newman and collaborators have found a partial
differential equation for the complex function of six real variables
$\Lambda(x;\zeta)$, which implies the existence of the metric $g$.  Once
such equation is solved, the explicit form
of $g$ can be found from $\Lambda(x;\zeta)$ by derivation. The magic is
then that the equation for the conformal factor that implies that $g$ solves
the Einstein equations is just a differential equation (an ODE, not a
PDE !).

This strange way of looking at GR reveals a new side of the theory, and
emphasizes the peculiarity of general relativity.  As Ted Newman puts it, a
conventional field theory describes propagation along characteristics of
certain operators; but general relativity is a theory determining its own
characteristics.  He suggests that this result indicates that GR has such a
peculiar structure that no known form of quantum mechanics
could be merged with it.   Certainly this beautiful result indicates that
general relativity is still capable of surprising us, and emphasizes how
deeply GR is a relational theory: the theory can be viewed as
a theory of the relative position --the position with respect to each
others-- of overlaying foliations.

{\bf References}

\noindent- S Frittelli, C Kozameh, ET Newman:
``On the dynamics of characteristic
surfaces", to appear on J. Math. Phys., Nov 1995.

\noindent- S Frittelli, C Kozameh, ET Newman:
``GR via characteristic surfaces",
submitted to J. Math. Phys.,1995.

\noindent- S Frittelli, C Kozameh, ET Newman:
``Linearized Einstein theory via null
surfaces", submitted J. Math. Phys., 1995.

\noindent- S Fittelli, C Kozameh, ET Newman:
``Lorentzian metrics from characteristic
surfaces", submitted to J. Math. Phys.1995.

\vfill\eject
\centerline{\bf ITP program solicitation}
\medskip
\centerline{James B. Hartle, University of California at Santa Barbara}
\centerline{hartle@itp.ucsb.edu}
\bigskip

The Institute for Theoretical Physics is an NSF funded institute located
on the Santa Barbara campus of the University of California. Its purpose
is to foster the progress of theoretical physics, especially in areas
where the traditional subfields overlap. It does this chiefly by
organizing 4-6 month research programs in which groups of scientists
in residence at the ITP explore specific problem areas. \footnote{*}{
More information about the ITP can be found from its homepage:\hfil\break
http://www.itp.ucsb.edu.}

Input from the
scientific community is central to determining the programs the ITP runs.
I am writing to encourage relativists to submit proposals for research
programs in areas where you believe that ITP activities can make
significant contributions. The main emphasis is on full, multi-month
programs, but ideas for workshops or mini-programs lasting a few
weeks or longer are also welcome. Criteria for the selection of programs
include intellectual significance, timeliness, suitability for the ITP,
experimental or observational significance, and availability of
outstanding participants.

The cycle for selecting programs for 1997-1998 year will begin with the
Advisory Board's steering committee meeting at the very end of this
September. At this early stage of program development, proposals
need not be elaborate --- a title, a paragraph or two explaining the idea,
and some suggestions for organizers and program participants will be
sufficient. For maximum utility suggestions should be made several weeks
in advance of the end of September so that they can be distributed to
the steering committee. Program suggestions can be sent to
Prof. James Langer, Director, Institute of Theoretical Physics,
University of California, Santa Barbara, CA 93106, or to langer@itp.ucsb.edu,
or to me at the same address, or at hartle@itp.ucsb.edu.

Please do not hesitate to get in touch with me at 805-893-2725 or
hartle@itp.ucsb.edu if you have questions.

\vfill\eject

\centerline{\bf 7th Gregynog workshop in general relativity}
\medskip
\centerline{Miguel Alcubierre, University of Wales at Cardiff}
\centerline{mam@astronomy.cardiff.ac.uk}
\centerline{http://www.astro.cf.ac.uk/groups/relativity}
\bigskip
\bigskip

The 7th Gregynog Workshop on General Relativity focused on Numerical
Relativity and took place from Monday August the 21th to Thursday
August the 24th, 1995.  It was attended by some 40 researchers from the
following countries:  Austria, Brazil, France, Germany, Greece, India,
Japan, Mexico, Russia, South Africa, Spain, the United Kingdom and the
United States.

Gregynog is a beautiful stately home surrounded by woods in the middle
of Wales.  Its relaxed atmosphere is ideal for workshops of this type,
since it allows for extensive interaction among the participants after
the conference sessions.

The programme for the workshop included many different aspects of
numerical relativity: critical phenomena in gravitational collapse,
relativistic stars, black-hole collisions, quasi-normal modes,
Cauchy-Characteristic matching, connection approaches to numerical
relativity, cosmology, numerical and visualization techniques, etc.

Since it would be impossible to summarize all the talks in this space,
I will concentrate my attention on those subjects that I consider to
have been the highlights of the meeting:

\noindent
1) Critical phenomena and self-similarity in gravitational collapse.

   Charles R. Evans presented a review of this exciting field.  Other
   related talks where given by R. Hamade, C. Gundlach and T.P. Singh.

   It is clear that the field has greatly developed since Choptuik's
   original discovery of critical phenomena in scalar field collapse.
   Many other physical systems have now been shown to have a similar
   behavior (gravitational wave collapse, radiation field collapse,
   relativistic fluid collapse),  and studies have been performed both
   in spherical and axial symmetry.

   The phenomena seems to be relatively well understood numerically.
   Great progress has also been made analytically making use of
   renormalisation group techniques.

   The critical exponents found for near critical collapse seem to be
   truly universal for a given physical system.  However, they are now
   known not to be universal across different physical systems.

\noindent 2) Black-hole collisions.

   The head-on collision of two black holes seems to be very well
   understood.  The numerical work of the NCSA-Washington University
   collaboration has provided us with beautiful pictures of the
   evolution of both apparent and event horizons.  Of particular
   interest has been the study of the dynamical oscillations of the
   horizon geometry, as well as the caustic structure of the horizon
   merger.  This calculations where originally done in 2D and have
   recently been repeated in 3D showing a very close agreement.

   One of the most important results to emerge form the study of this
   problem has been the remarkable agreement between full numerical
   evolutions and linearized approximations.  The work of J. Pullin and
   R. Price has shown that the linear approximation works surprisingly
   well in a strong field regime.

\noindent 3) Cauchy-characteristic matching.

   A technique that finally seems to be giving its first fruits is that
   of Cauchy-characteristic matching.  The work presented at the
   workshop by R. d'Inverno shows that the technique works very well
   for 1D problems.  Its application to more than one dimension is also
   being developed, as was shown to us in the work of N. Bishop.

   This technique, with its promise of eliminating arbitrary boundary
   conditions and allowing calculations to go out all the way to Scri+,
   will be essential for numerical relativity in the future.

\noindent 4) Hyperbolic relativity.

   Over the last few years the need to understand the characteristic
   structure of the evolution equations of general relativity has
   become apparent.  Writing the evolution equations in a fully
   hyperbolic form is a fundamental step both from the analytical and
   numerical points of view.

   The works presented by C. Bona and J. Masso have shown how it is
   indeed possible to write the evolution equations in hyperbolic form
   in a way that allows the use of very powerful numerical techniques.

   Hyperbolic relativity is one of the most fundamental developments of
   the last few years and will surely have a great impact in numerical
   relativity in the near future.

\noindent 5) 3D numerical relativity.

   During the workshop it became clear that 3D numerical relativity is
   finally here.  There are now several groups working on 3D codes on
   both sides of the Atlantic.

\vfill\eject
\centerline{\bf 3rd Annual Penn State Conference:}
\centerline{\bf Astrophysical Sources of Gravitational Waves}
\medskip
\centerline{Curt Cutler, The Pennsylvania State University}
\centerline{curt@phys.psu.edu}
\bigskip

   This conference was organized by C. Cutler and K. Thorne, and took
place at Penn State during July 7--11, 1995.  It was the third gravity
conference in an annual series at Penn State, the first two having
focused on numerical relativity and quantum geometry, respectively.
The Conference was divided into two parts: 2 days of invited and
contributed talks intended to give a broad overview of the field to
people who might be new to it, followed by a 2-day workshop concerned
more narrowly with coalescing neutron star and black hole binaries.
This workshop was the follow-up to a similar workshop held at Caltech in
Jan., 1994.  Roughly 85 people attended part I of the Conference, with
the great majority also staying on for part II, the workshop.  In part
I, the invited speakers and their titles were:

\noindent Albert Lazzarini   ``LIGO and the Users Community''

\noindent Mark Bocko         ``Overview of TIGA and other Spherical,
                           Resonant Bar Detectors''

\noindent Vincent Loriette   ``VIRGO Status Report''

\noindent Adam Burrows       ``Supernovae''

\noindent Ewald Mueller      ``The Gravitational Wave Signature
                       of Rotational Core Collapse''

\noindent Dong Lai           ``Hydrodynamical Processes in Newborn,

\noindent Spinning NS's and in Merging
                          NS binaries, and their Gravitational
\noindent Wave Signatures''

\noindent Bruce Allen        ``Gravitational Waves from the Early Universe''

\noindent Peter Bender       ``LISA Overview and Target Sources for LISA''

\noindent Clifford Will      ``Coalescing Compact Binaries''

\noindent Misao Sasaki       ``Black Hole Perturbation
Approach to Gravitational
                         Radiation from Compact Binary Systems''

\noindent Bernard Schutz     ``Gravitational Waves from Rotating
Neutron Stars''

There were also contributed talks by G. Quinlan, F. Ryan,
S. Chakrabarti, W. Suen, E. Seidel, and K. New, and D. Nicholson
contributed an especially stimulating talk on non-linear filtering and
adaptive line enhancement.

   The somewhat unusual format chosen for the workshop seemed to work
rather well. Ten sets of scientific issues relating to binary
coalescence were identified, and then for each set one or two experts
in that area were chosen to moderate a (roughly) one-hour discussion.
The hour generally began with a short presentation by the expert(s)
that outlined the basic problems, which was followed by a few 5-minute
talks in which people described their latest (and often unpublished)
results in these areas, and ended with a group discussion.  The group
discussions were often lively. One resulted in a bet between R. Price
and K. Thorne on the question of whether or not, in the final
coalescence of two spinning black holes with non-aligned spin axes,
most of the gravitational wave energy emitted after the inspiral phase
will essentially take the form of a simple, quasi-normal mode
ringdown.  This question was related to an extended discussion on the
proper role of the BH Grand Challenge Project vis-a-vis LIGO.

    Pleasantly, it was clear that a lot of progress had been made
since the previous workshop, especially in the areas of PN
calculations of inspiral templates (by the groups of Will and Wiseman
and Blanchet, Damour, and Iyer), and in estimating the computing power
required to do a real-time time search of LIGO data for coalescing
binaries ( $<$ 300 Megaflops, according to B. Owen). Especially
interesting was the re-formulation by Will and Wiseman of the
Epstein-Wagoner approach to doing PN calculations, in such a way as to
remove, at all orders, the infinities that had muddied the
Epstein-Wagoner approach.

    The meeting ended with a general discussion in which it was agreed
that a formal organization of theorists interested in LIGO-related
research should be created, in order to represent their concerns to
LIGO Management, to be a body to which LIGO Management can turn for
theoretical input, and to sponsor more meetings.  A self-destructing
committee was appointed to write a charter and oversee the first
election of a Steering Committee for that organization.

\vfill\eject
\parskip=3pt
\centerline{\bf General news from GR14}
\medskip
\centerline{Abhay Ashtekar, The Pennsylvania State University}
\centerline{ashtekar@phys.psu.edu}
\medskip

$\bullet${\bf Elections} During the general assembly
the
following election results were announced. J\"urgen Ehlers,
director of the newly founded Max Planck Institute for Gravitational
Physics in Potsdam, Germany, is the new President.  He succeeds Sir
Roger Penrose of Oxford and Penn State Universities. According to the
rules of the Society, Professor Penrose will now serve as the
Vice-President. Professor Malcolm MacCallum of Queen Mary and
Westfield College, London was elected as the Secretary.  Professors
Ehlers, Penrose and MacCallum constitute the executive committee of
the Society. Professor Clifford Will of Washington University at
St. Luis was elected as the new U.S. representative on the
International Committee of the Society. He will serve a nine year
term.

$\bullet${\bf GR15}
The next tri-annual international conference on General
Relativity and Gravitation will take place in Poona, India in December
of 1997. It will be hosted by the Inter-University Center for Astronomy
and Astrophysics (IUCAA).  The conference was moved to December
because June and July fall in the monsoon season in India and because
December-January period is the best for travel within the country.
1997 was chosen (as opposed to 1998) to avoid conflict with the Texas
symposium.  Professor E. T. Newman of University of` Pittsburgh will
serve as the Chair of the Scientific committee. If you have general
suggestions for the format of the scientific program --e.g. on
distribution of topics, number of workshops, etc-- please let him know
soon. Planning for the conference will begin already this fall since
the committee has half a year less than usual to come up with the
program.

$\bullet${\bf Xanthopoulos Prize}
{}From now on, the Basilis Xanthopoulos Prize for General
Relativity and Cosmology will be presented during the International
Conferences on General Relativity and Gravitation. The prize is given
for outstanding original work in our field to researchers less than
forty years of age. Preference is given to theoretical work. It
carries a monetary award of approximately \$ 10,000 and a citation.
The endowment comes from the Foundation for Research and Technology --
Hellas (FORTH) which is based in Heraklion,Crete, where Basilis
Xanthopoulos taught for many years before his untimely death.  The
winners are selected by an international committee of relativists.

This year's Xanthopoulos prize was awarded to Professor Carlo Rovelli
of the University of Pittsburgh during the general assembly of the
Society.  Professor Sotirios Persidis opened the ceremony by recalling
how the prize originated. (For details, see Matters of Gravity, number
3). Professor Penrose then presented the prize. The citation read:
{\it Forth Foundation and the Selection Committee is pleased to
present the 1995 Xanthopoulos award to Professor Carlo Rovelli of the
University of Pittsburgh for his wide ranging contributions to
classical and quantum gravity, in particular for his stimulating
papers on the issue of physical observables in diffeomorphism
invariant theories and his pioneering ideas in the development of the
loop representation in quantum general relativity}. The ceremony
concluded with an acceptance speech by Professor Rovelli. The full
text of this acceptance speech as well as the opening remarks by
Professor Persidis will  appear in the proceedings of the conference.

\vfill\eject
\centerline{\bf Canadian General Relativity Conference}
\medskip
\centerline{Jack Gegenberg, University of New Brunswick}
\centerline{lenin@math.unb.edu}
\bigskip

Black holes seemed to occupy the center of discourse, much as they
(conjecturally) occupy centers of certain galaxies, at the {\it Sixth
Canadian General Relativity and Relativistic Astrophysics Conference}
which took place May 25-27 at the University of New Brunswick in
Fredericton, NB.

Jim Isenberg and Charles Torre discussed mathematical foundations of
classical and quantum gravity.  Jim reported on the work of he and his
group on solutions of the constraints when the mean curvature is not
constant.  Charles reported on the search (so far unsuccessful) for
generalized symmetries in GR.

John Friedman discussed problems that appear when
 formulating quantum field theory in
non-globally hyperbolic geometries-i.e. in spacetimes where time travel is
allowed!

Alan Coley and Michael West reported on topics in cosmology.  Alan discussed
self-similar models; while Michael built a persuasive case for a strong role
of black holes in the formation of the galactic structures.  This provided
the transition to a number of talks on various problems in black hole physics.

Werner Israel and Bob Wald provided surveys of, respectively, the classical
and semiclassical physics of black hole interiors, centered around the
phenomena of mass inflation; and of the laws of black hole mechanics in
arbitrary gravity theories, using his ``Noether charge" formalism.

Cliff Burgess and Gabor Kunstatter reported on their respective work in
stringy black holes and in 1+1 dimensional dilaton-gravity models.  In
particular, Cliff discussed the role of superstring duality transformations
in black hole physics; while Gabor discussed the interpretation of observables
in generic dilaton-gravity theories.

Finally, Eric Poisson and Jorge Pullin reported on the progress of
their work on numerical relativity.  Eric very kindly put together a
plenary talk to replace another speaker who was unable to present.  He
surveyed the issues involved in the production of gravity waves from
colliding binaries.  Jorge presented a bravura multi-media display of
a numerical study of the linearized theory of colliding black holes.

We heard contributed talks from a number of gravity-oids, many of which
are grad students and postdocs from all over North America and beyond.
The plenary and contributed talks will be published by the Fields Institute
for Research in Mathematical Sciences.
We wish to thank
them, as well as the Natural Sciences and Engineering Research Council of
Canada, the Canadian Institute for Theoretical Astrophysics and the
Universities of New Brunswick and
Moncton for partial support.  We also thank the many lobsters that very
graciously allowed themselves to be eaten by us at our conference banquet.

\end